\documentclass[acmsmall,screen]{acmart}
\usepackage{tcolorbox}
\usepackage[ruled,vlined]{algorithm2e}
\AtBeginDocument{%
  }

\setcopyright{acmcopyright}
\copyrightyear{2018}
\acmYear{2018}
\acmDOI{XXXXXXX.XXXXXXX}

\acmBooktitle{Proceedings of the 32nd ACM Symposium on the Foundations of Software Engineering (FSE '24), November 15--19, 2024, Porto de Galinhas, Brazil}

\acmPrice{15.00}
\acmISBN{978-1-4503-XXXX-X/18/06}

\usepackage{multirow}
\usepackage{soul}

\begin{document}

\title{From Bias To Improved Prompts: A Case Study of Bias Mitigation of Clone Detection Models}


\author{QiHong Chen}
\email{chenqh@uci.edu}
\affiliation{%
  \institution{University of California, Irvine}
  \city{Irvine}
  \state{California}
  \country{USA}
}

\author{Lianghao Jiang}
\email{lianghaj@uci.edu}
\affiliation{%
  \institution{University of California, Irvine}
  \city{Irvine}
  \state{California}
  \country{USA}
}

\author{Iftekhar Ahmed}
\email{iftekha@uci.edu}
\affiliation{%
  \institution{University of California, Irvine}
  \city{Irvine}
  \state{California}
  \country{USA}
}

\renewcommand{\shortauthors}{Trovato et al.}

\begin{abstract}

The issue of clone code has persisted in software engineering, primarily because developers often copy and paste code segments. This common practice has elevated the importance of clone code detection, garnering attention from both software engineering researchers and industry professionals. Their collective concern arises from the potential negative impacts that clone code can have on software quality. The emergence of powerful Generative Large Language Models (LLMs) like ChatGPT has exacerbated the clone code problem. These advanced models possess code generation capabilities that can inadvertently create code clones. As a result, the need to detect clone code has become more critical than ever before. In this study, we assess the suitability of LLMs for clone code detection. Our results demonstrate that the Palm model achieved a high F1 score of 89.30 for the avatar dataset and 86.41 for the poolC dataset. A known issue with LLMs is their susceptibility to prompt bias, where the performance of these models fluctuates based on the input prompt provided. In our research, we delve deeper into the reasons behind these fluctuations and propose a framework to mitigate prompt bias for clone detection. Our analysis identifies eight distinct categories of prompt bias, and our devised approach leveraging these biases yields a significant improvement of up to 10.81\% in the F1 score. These findings underscore the substantial impact of prompt bias on the performance of LLMs and highlight the potential for leveraging model errors to alleviate this bias.

\end{abstract}




\maketitle

\section{Introduction}
\label{sec:intro}



Code clones are source code fragments that exhibit either syntactic or semantic similarity~\cite{krinke2001identifying, baxter1998clone, ducasse1999language,komondoor2001using}. They are prevalent in traditional software development practices, often arising from the practice of copying and pasting source code snippets found on online platforms like Stack Overflow~\cite{ragkhitwetsagul2019toxic, zhang2012cloning} and duplicated repositories on platform such as GitHub~\cite{lopes2017dejavu}. However, clone codes have the potential to detrimentally affect the quality of traditional software, leading to issues in software maintenance and debugging~\cite{zhang2021survey}. Moreover, from a software system security perspective, clone codes could lead to vulnerability propagation if a vulnerable code fragment is cloned~\cite{li2016clorifi}. Consequently, clone detection has emerged as a crucial field in software engineering, tasked with predicting whether two code snippets are indeed clone codes. This area of research has gained significant traction due to its practical applications, including code reusability enhancement, aspect mining, plagiarism detection, copyright infringement investigation, code optimization, software evolution analysis, code quality assessment, bug detection, virus detection, and improvement of code maintainability~\cite{roy2009comparison}. 

In the realm of traditional software, clone codes have been a prevalent occurrence. However, the advent of cutting-edge large language models (LLMs), exemplified by ChatGPT, has significantly amplified the prevalence of clone codes in both traditional and machine-learning-based software. This amplification can be attributed to LLMs' exceptional capabilities in source code generation and human-language comprehension. These remarkable features have captivated users across a spectrum of coding proficiency levels, empowering them to generate source code from natural language requirements, incomplete code fragments, and even code written in other programming languages~\cite{liu2023refining, khoury2023secure, liu2023improving}. Consequently, this exacerbates the issue of clone codes because LLMs are bound by the source code they have encountered during their training phase. When users prompt LLMs with similar requirements, they generate similar source code, leading to a proliferation of similar code instances. Moreover, as reported by Liu et al.~\cite{liu2023refining}, the quality of LLMs-generated source code heavily relies on the difficulty of the requirement, program size and the programming language. Thus, when these LLM-generated source codes are incorporated into machine-learning-based software, its quality can be significantly compromised. Furthermore, Allamanis et al.~\cite{allamanis2019adverse} underscore the impact of clone code on machine learning models. They highlight that, in certain scenarios, a machine learning model's performance can experience inflation of up to 100\% when evaluated on datasets containing duplicated code, as opposed to its performance on datasets where duplicated code has been eliminated. Therefore, with the ascendancy of LLMs, the importance of clone detection has been elevated to a paramount position in software development domain.

Over the years, researchers have devised a multitude of methodologies, encompassing text-based, lexical-based, tree-based, metric-based, semantic-based, hybrid-based, and machine learning-based approaches for clone detection~\cite{roy2009comparison,saini2018code, ain2019systematic}. Among these approaches, the machine learning paradigm exhibits a particularly strong alignment with clone detection task due to its heightened emphasis on comprehending the nuanced meaning embedded within source code. Numerous machine-learning approaches have been proposed for clone detection. For instance, Fang et al.\cite{Fang2020} employed a deep neural network (DNN) to identify function clones, capturing syntax and semantic information through abstract syntax trees (ASTs) and control flow graphs (CFGs). Wang et al.~\cite{wang2020detecting} proposed an approach combining graph neural networks and control/data flow information that outperformed state-of-the-art techniques using the AST subtree-based approach for clone detection. An essential aspect of clone code detection is the ability to comprehend the source code, and research has shown that Large Language Models (LLMs) have robust code comprehension capabilities. This makes LLMs a perfect candidate for clone detection. However, current research has not investigated the benefit of using LLms for clone detection and we aim to fill this gap in this research.


While LLMs possess remarkable source comprehension abilities, they suffer from issues like prompt bias~\cite{zhou2022large} and hallucination. Prior research has highlighted that LLMs' outputs fluctuate substantially when prompts are slightly modified. Zhou et al.~\cite{zhou2022large} introduced an approach to uncover improved prompts by incorporating descriptive context alongside desired input/output, thereby instructing the GPT model to generate the desired prompt. However, this method has limitations as it does not address the root cause of the output variations. We assert that the fundamental source of prompt bias can be traced back to the errors made by the model. By comprehending these errors, we can refine the prompt to optimize the performance of LLMs. Consequently, we seek to bridge this gap by applying LLMs to clone detection using an innovative approach to mitigate the prompt bias inherent in LLMs.


In this study, our objective is to employ LLMs for both within-language and cross-language clone detection tasks. We have made the deliberate choice not to confine our investigation solely to within-language clone detection. This choice is motivated by the recognition that cross-language clone codes represent a particularly challenging category, being semantically similar yet textually distinct. We contend that clone codes of this category hold greater relevance in real-world scenarios. Furthermore, by demonstrating the effective capability of LLMs in identifying both within and cross-language clone codes, we aim to underscore the pivotal role of LLMs in shaping future trends in clone detection research. To achieve this goal, we conduct experiments to answer the following research questions:
\begin{itemize}
    \item RQ1: In the context of within-language and cross-language clone detection tasks, what are the prompt bias mistake categories?
    \item RQ2: How prevalent are the mistake categories and what is the most frequent category?
    \item RQ3: How can we harness these categories to improve the model performance?
\end{itemize}

\noindent Specifically, this paper makes the following contributions:
\begin{enumerate}
    \item This study represents the first exploration of LLMs within the domain of clone detection.
    \item  This is the first study that marks the pioneering effort to delve into the prompt bias of LLMs from the model's error perspective. 
    \item We introduce a comprehensive framework meticulously crafted to augment the performance of GPT models in both within-language and cross-language clone detection tasks.

\end{enumerate}
\label{sec:intro}
\section{Background \& Related Works}

In this section, we commence by offering an overview of the historical evolution of research on clone detection. 

\subsection{The Evolution of Code Clone Detection}
Clone code is a pair of similar source codes that are either identical or appear like one another within a software system. Roy et al.~\cite{roy2007survey} outline the fundamental types of clone code as following:\\\textit{\textbf{Exact Clones (Type 1)}}: Identifical code segments except for changes in comments, layouts and whitespaces.\\\textit{\textbf{Renamed Clones (Type 2)}}: Code segments which are syntactically or structurally similar other than changes in comments, identifiers, types, literals, and layouts. These clones are also called parameterized clones.\\\textit{\textbf{Near Miss Clones (Type 3)}}: Copied pieces with further modication such as addition or removal of statements and changes in whitespaces, identifiers, layouts, comments, and types but outcomes are similar. These clones are also known as gapped clones.\\\textit{\textbf{Semantic Clones (Type 4)}}: More than one code segments that are functionally similar but implemented by different syntactic variants.

In the realm of clone detection research, various methodologies have been devised, encompassing text-based, lexical-based, tree-based, metric-based, semantic-based, hybrid-based, and machine learning-based approaches. However, it is widely recognized among researchers that a mere focus on syntactic similarity falls short of capturing the essence of code functionality and the crux of effective clone detection lies in comprehending the functional aspects of source code. An innovative breakthrough in this pursuit was introduced by Yu et al.~\cite{YU2023107130}, who employed a semantics learning method utilizing Control Flow Graphs (CFG) and Program Dependency Graphs (PDG) to delve into deeper semantic intricacies. Subsequent research endeavors expanded upon this approach by creating additional source code representations. For instance, CCGraph~\cite{7917935} harnessed the power of Program Dependency Graphs (PDGs) in conjunction with near graph-matching algorithms to detect code clones. DeepSim~\cite{zhao2018deepsim} took this a step further by encoding control and data flow as semantic matrices, offering a more comprehensive representation of code semantics. These machine learning-based hybrid approaches outperformed approaches that relied on techniques using the AST subtree-based approach for clone detection. Despite the divergence in these approaches regarding the representations they employ, their common objective remains consistent: aiding BERT models in better grasping source code semantics.

\subsection{Large Language Models Use Cases}

In contrast to BERT models, LLMs such as ChatGPT, whose behaviors are guided by prompts, exhibit unparalleled capabilities across a range of software engineering research areas. These include machine translation~\cite{chowdhery2022palm, hendy2023good, radford2019language}, content summarization~\cite{radford2019language, feng2020codebert, goyal2022news}, sentiment analysis~\cite{xu2020dombert, zhang2020sentiment}, and query resolution~\cite{radford2019language}. LLMs have acquired these exceptional skills through extensive training on vast textual datasets and innovative training methodologies, such as reinforcement learning from human feedback (RLHF)~~\cite{ouyang2022training}. These approaches enable LLMs to grasp intricate linguistic patterns and structures. Given their capabilities, prior research has primarily focused on integrating LLMs into various domains, including conversational agents like ChatGPT~\cite{chatgpt} and AI-assisted drawing, exemplified by Midjourney~\cite{midjourney}. Remarkably, the source code comprehension ability of LLMs aligns with the insights of clone detection researchers, although there has been limited exploration into the suitability of LLMs for clone detection tasks. In our study, we address this gap and empirically assess the performance of LLMs in the context of clone detection.

\subsection{Large Language Models Limitations}

While LLMs have demonstrated impressive capabilities across various domains, they are not without their limitations. Previous research has highlighted certain shortcomings inherent to machine learning models, such as dataset biases, encompassing issues like class imbalance and feature bias~\cite{gesi2021empirical, aggarwal2019black, biswas2020machine, chakraborty2020making, chakraborty2019software, harel2020neuron, tian2020testing}. Additionally, LLMs are prone to a phenomenon known as "hallucination," where they generate text that seems logically coherent and contextually relevant but may contain factual inaccuracies or entirely fabricated information~\cite{alkaissi2023artificial}. To address this hallucination issue, various techniques have been developed, such as self-consistency~\cite{wang2022self} and chain of thought~\cite{wei2022chain}, which aim to guide GPT models in producing reliable outputs by fostering internally consistent reasoning. Another significant challenge faced by LLMs is "prompt bias," which occurs when slight modifications to the input prompt result in substantially different model outputs. Zhou et al.~\cite{zhou2022large} attempted to tackle prompt bias by introducing an approach that identifies improved prompts. They achieved this by providing descriptive context alongside the desired input/output, encouraging the GPT model to generate the desired prompt. However, this approach primarily focuses on mitigating prompt bias by generating better prompts from the model's perspective. However, their approach does not delve into the underlying causes. In our study, we aim to bridge this gap by conducting an in-depth analysis of the factors driving LLM behavior and using the gained insights to address prompt bias effectively in the domain of clone detention, which, to the best of our knowledge, is the first of its kind.

\section{Experimental Setup}
\label{sec:method}

\begin{figure}[t]
    \centering
    \includegraphics[width=.8\columnwidth]{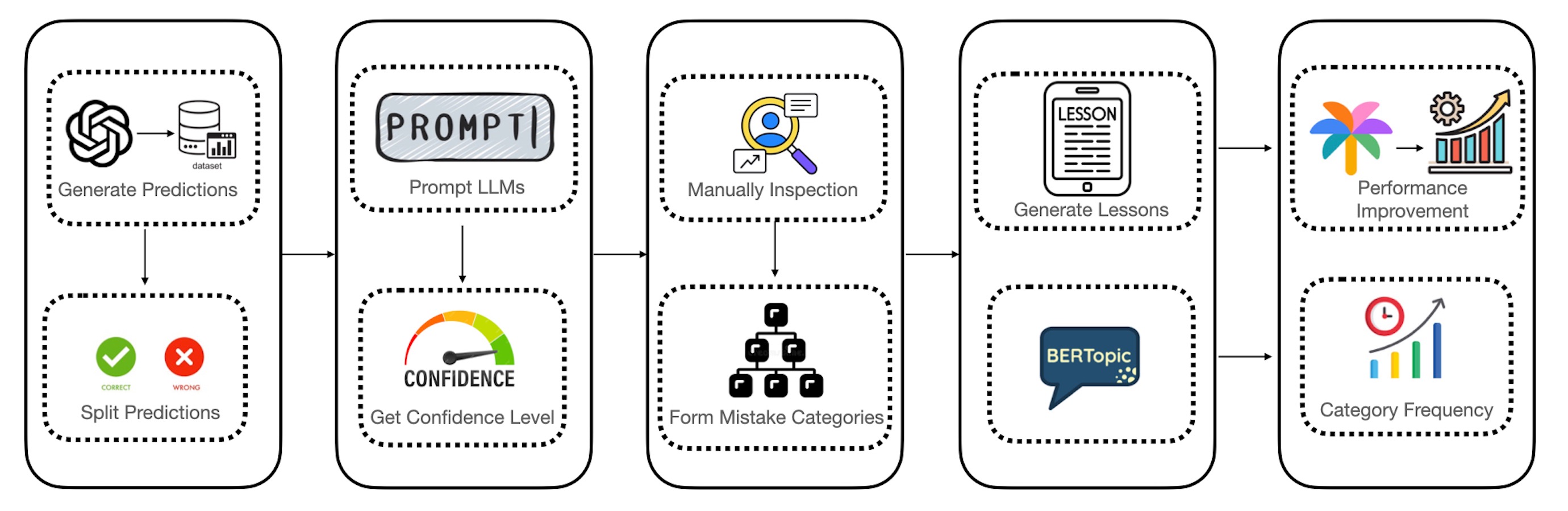}
    \caption{Overall Approach}
    \label{fig:RQ_Overview}
\end{figure}

In this section, we describe our process to answer each research question(RQs): (A) We choose the model for our later analysis. (B) We generate prompt bias mistake categories based on prompted model mistakes. (C) We calculate the prevalence of each prompt bias mistake category. (D) We propose a method to improve model's performances using the prompt bias lessons. (E) We evaluate the proposed method. An outline of our methodology is presented in Figure~\ref{fig:RQ_Overview}.

\subsection{Model Selection}

In this section, we outline our model selection process, covering dataset specifics, candidate model information, and the procedure for identifying the top-performing model.

\subsubsection{Datasets Introduction:} The primary goal of this paper is to investigate LLMs performance on clone detection for both within and cross-language scenarios. Among many available datasets, we picked the poolC dataset~\cite{poolC} for within-language clone detection and created an avatar dataset for cross-language clone detection based on a language translation project named avatar~\cite{avatar}. 

We decided to use poolC datset~\cite{poolC} since it has been extensively used in clone-related research. We opted not to use the well-known BigCloneBench (BCB) dataset due to issues related to unbalanced data labels and ground truth quality ~\cite{sonnekalb2022generalizability,krinke2022bigclonebench}. The avatar dataset originally served as a code translation dataset~\cite{avatar}, comprising code snippets with identical functionality but composed in distinct programming languages. We leveraged this dataset and transformed it into a cross-language clone detection dataset. The original avatar dataset comprised three JSONL files, where each line within the file represented a dictionary. The dictionary structure was as follows: \{"java": Java code snippet, "python": Python code snippet, idx: idx value\}. The "idx" value denoted the theme of the code snippets, and Java and Python code snippets were considered to share the same functionality ~\textit{iff} they possessed the same "idx" value. Given the consistent format across all three JSONL files, we consolidated them into a single JSONL file named "combined.jsonl," containing 9,515 lines. Subsequently, we constructed our avatar dataset from "combined.jsonl," comprising a "data.jsonl" file, a training dataset name "train.txt", a validation dataset "valid.txt", and a testing dataset named "test.txt". A comprehensive overview of these datasets can be found in Table ~\ref{tab:datasets}. 

Notably, the code snippets in the poolC and avatar datasets include comments. While many non-machine learning-based approaches tend to discard comments as they are often seen as non-essential to the functionality of source code, we intentionally chose to retain comments. Our reasoning is that the effectiveness of machine learning models relies on their ability to understand source code comprehensively. Comments, which serve as explanations of the code written by developers, have the potential to assist models in better comprehending the code. We created two versions for both the avatar and poolC datasets: one with comments and one without comments.

\begin{table}[t]
\footnotesize
\centering
\caption{Details of Datasets}
\label{tab:datasets}
\resizebox{\columnwidth}{!}{%
\begin{tabular}{|c|c|c|c|c|}
\hline
\textbf{Dataset}        &\textbf{Clone Detection Type}      & \textbf{Dataset Size} & \textbf{Sampled Size} & \textbf{Language}               \\ \hline
Avatar with comment     & Cross Language                    & 1.7k                  & 1.7K                  & Java, Python      \\ \hline
Avatar without comment  & Cross Language                    & 1.7k                  & 1.7K                  & Java, Python           \\ \hline    
PoolC with comment       & Within Language                   & 6,710k                 & 16K                   & Python                   \\ \hline 
PoolC without comment    & Within Language                   & 6,710k                 & 16K                   & Python                   \\ \hline    
\end{tabular}%
}
\end{table}

\subsubsection{Candidate Models Introduction:} Among all available deep learning models and LLMs, we picked CodeT5~\cite{wang2021codet5}, a BERT model with 220 million parameters, Turbo-3.5~\cite{openai_models} and Text-002 models~\cite{openai_models}, both GPT models developed by OpenAI with 175 billion parameters, and PaLM model~\cite{chowdhery2022palm}, another GPT model by Google boasting a substantial 540 billion parameter size. We included models from both Bidirectional Encoder Representations from Transformers (BERT) and Generative Pre-trained Transformer (GPT) model families because of the absence of sufficient evidence indicating the superior performance of GPT models over BERT models in the context of clone detection task. Since Google and OpenAI do not offer direct access to their model weights, we interact with these models using prompts. In contrast, codeT5 model had been pre-trained, and we subsequently fine-tuned it on the dataset to maximize the performance.

\subsubsection{Selecting Model:}\label{subsubsec: Selecting Model} To choose one model for further analysis, we ran every model on every dataset and picked the top-performing model. Given the substantial size of each dataset and the considerable amount of time and budget required to evaluate models on the entire dataset, we opted to establish a benchmark dataset comprising 200 instances. For every dataset, we randomly sampled 200 instances from the validation set. To mitigate any potential adverse effects stemming from class imbalance, we took great care to maintain class balance within the benchmark dataset, ensuring an equal distribution of clone and non-clone instances. Next, we evaluated every model on every crafted benchmark dataset and recorded the model performances in terms of precision, recall, accuracy and F1 score. For CodeT5, we finetuned it on the training set and evaluate it on the test set. For GPT model, we employed the default prompt defined with a temperature setting of 0 to evaluate the model performance on the test set. We configured the temperature to be 0, ensuring that the model's output is deterministic and our results are reproducible. To construct the default prompt, we adopted Zhou et al's~\cite{zhou2022large} approach and formulated our default prompt using GPT-4. Specifically, we feed GPT-4 model with several exemplary input/output instances and request it to generate the default prompt. Note that we deliberately include programming languages of the code snippets in the prompt because we posit that programming language information could assist the model in better understanding the code snippet~\cite{liu2023refining}. The default prompt we used is shown in Figure ~\ref{pic:default_prompt}. Lastly, we picked the top-performing model in terms of F1 score.

\begin{figure}[h]
\begin{tcolorbox}
\textit{You are a helpful assistant that takes two pieces of code snippets and detects whether these two pieces of code snippets are clone code or not. The first code snippet is written in <language1>, and the second code snippet is written in <language2>. Here are the two code snippets: Code Snippet1: <code1>. Code Snippet2: <code2>. If you think any part of the two code snippets contains clones, answer Yes, otherwise answer No. You must answer either Yes or No. Your Response:[INSERT].}
\end{tcolorbox}
\caption{Default Prompt}
\label{pic:default_prompt}
\end{figure}


\subsection{Prompt Bias Mistake Categories Generation}
In this section, we describe the procedure for generating prompt bias categories. Specifically, this includes generating model predictions, getting model confidence levels, and generating prompt bias mistake categories.

\subsubsection{Generate Model Predictions:} Due to the substantial size of the poolC dataset, we employed a sampling approach~\cite{dell2002sample} using a sampling size calculator~\cite{monkeys} with a 99\% confidence level and a 1\% margin of error to construct the 16k size poolC with comment dataset and 16k size poolC without comment dataset. Next, we ran the selected model on every dataset and recorded the model performance. For BERT model, we finetune it on the training set and, subsequently, apply it to the test set to evaluate its performance. For the GPT model, we employed the default prompt defined in~\ref{subsubsec: Selecting Model} with a temperature setting of 0 to evaluate the model performance on the test set. 

\subsubsection{Get Model Confidence Level:}\label{subsec:confidence} We then divided the model's predictions into correct and incorrect prediction groups. The correct prediction group consists of instances where the prediction is the same as the ground truth label and the incorrect prediction group consists of instances where the prediction is different from the ground truth label. To study the mistakes of the model, we focused on the incorrect prediction group. To mitigate the negative effect caused by hallucination~\cite{lee2023mathematical}, we first extracted the confidence level of the model regarding its prediction using the prompt shown earlier where we explicitly asked \textit{Tell me how much confidence you have for your answer scale from 0 to 100.} The prompt we used to get confidence level of the model is shown in Figure~\ref{pic:confidence_prompt}.

\begin{figure}[h]
\begin{tcolorbox}
\textit{"You are a helpful assistant that takes two pieces of code snippets and detects whether these two pieces of code snippets are clone code or not. The first code snippet is written in <language1>, and the second code snippet is written in <language2>. Here are the two code snippets: Code Snippet1: <code1>. Code Snippet2: <code2>. If you think any part of the two code snippets contains clones, answer Yes; otherwise answer No. You must answer either Yes or No. Your Response:[INSERT]." I have documented your answer as <prediction>. Tell me how much confidence you have for your answer scale from 0 to 100.} 
\end{tcolorbox}
\caption{Confidence Extraction Prompt}
\label{pic:confidence_prompt}
\end{figure}

We employed this confidence level to indicate the reliability of model's responses. Therefore, we retained only those incorrect predictions where the model exhibited 80\% confidence in its responses, as these instances represented cases where the model made mistakes with a high degree of confidence. We call them reliable incorrect predictions. We picked 80\% as a cutoff point because, as we observed, the model's confidence level is either higher than 80\% (from 80\% to 100\%) indicating highly confident or 0\% indicating not confident. Therefore, picking 80\% as a cutoff point was a reasonable choice. We provide the statistical detail of selecting threshold value later in section~\ref{subsection: RQ2}.


\subsubsection{Prompt Bias Mistake Categories Generation:}
\label{subsubsec:Prompt Bias Mistake Categories Generation} To understand the reasons behind the model's incorrect predictions, we extracted the high-level rationale behind the incorrect predictions. The prompt we used is shown in Figure~\ref{pic:rational_extraction_prompt}. These high-level rationales encompass the model's thought process during the prediction phase. To understand the root cause behind those rationales, we randomly selected 100 examples and reviewed them. To mitigate human bias, two authors independently reviewed the 100 instances together with the code snippets meticulously, with the objective of summarizing the errors made by the model that are specifically related to the misunderstanding of the nature of the clone detection task~\cite{strauss1998basics}. Subsequently, the two authors  discussed the list of model mistakes that pertained to misinterpretations of the task and reached an inter-rater reliability Cohen's Kappa (k)~\cite{gwet2008computing} of 0.85, which indicates almost perfect agreement. These identified model errors were coined as "prompt bias mistake categories."


\begin{figure}[h]
\begin{tcolorbox}
"You are a helpful assistant that takes two pieces of code snippets and detects whether these two pieces of code snippets are clone code or not. The first code snippet is written in <language1>, and the second code snippet is written in <language2>. Here are the two code snippets: Code Snippet1: <code1>. Code Snippet2: <code2>. If you think any part of the two code snippets contains clones, answer Yes, otherwise answer No. You must answer either Yes or No. Your Response:[INSERT]." Your answer was <model\_prediction>. However, the correct answer is <label>. Your answer was wrong,  I want you to give me the reason why you made the decision in high level.
\end{tcolorbox}
\caption{Rational Extraction Prompt}
\label{pic:rational_extraction_prompt}
\end{figure}

\subsection{Calculating Prevalence of Prompt Bias Mistake Category:} As we have defined the prompt bias mistake categories, next we want to understand the prevalence of each category. We adopted a topic modeling tool named BERTopic~\cite{grootendorst2022bertopic} to calculate the frequency of each category. Our applied procedure was as follows:
\begin{enumerate}
\item \textbf{Generated Topics:} We adopted the methods mentioned above in subsection ~\ref{subsubsec:Prompt Bias Mistake Categories Generation} to generate rationales for every incorrect instance of every dataset. Those rationales are then fed into BERTopic as the input data. BERTopic is designed to be modular, meaning each algorithm phase may be customized. As such, ensuring the appropriate configuration at each step is of the utmost importance. The first phase in the BERTopic pipeline is calculating embedding vectors for the input documents. Various embedding models are supported in this step. Thus, our first variable that can significantly affect the final topics is the embedding model employed. We opted to use the default embedding model which is all-MiniLM-L6-v2. 

In the second phase, the dimensionality of the embedding vectors is reduced to counteract the curse of the dimensionality problem in the clustering phase. To handle large amounts of documents from poolC dataset, cuML's UMAP implementation can be used to enable GPU utilization in this phase. Although other dimensionality reduction algorithms could be used, we adopted UMAP as its superiority is discussed in previous work \cite{grootendorst2022bertopic,mcinnes2018umap}. CuML's UMAP offers a range of tunable parameters, but among them, two variables highlighted by BERTopic have a substantial impact on the algorithm: the \textbf{number of neighbors} and the \textbf{number of components}. The number of neighbors, which can be set between 2 and 100, plays a crucial role in determining the preservation of local data. Smaller values favor the retention of more localized data. Additionally, the number of components represents the dimensionality of the reduced vectors. Increasing the value of this variable reduces the impact of dimensionality reduction on clustering. In addition to the aforementioned hyperparameters, there is another crucial parameter to consider in cuML UMAP: the \textbf{minimum distance} between the reduced vectors. Ranging from 0 to 1, this parameter plays a pivotal role in the balance between global and local structure preservation. When set to smaller values, it leads to a more tightly clustered embedding space, where vectors in close proximity on the manifold are drawn even closer together. We initiated with a conservative minimum value of 5 for both hyperparameters. Subsequently, we incremented these values in steps of 5, gradually increasing them to 20. This enabled the exploration of larger values in cases where we observed an upward trend in the topic model's performance.

In the third phase, reduced vectors are clustered. Similar to previous phases, BERTopic supports customized clustering algorithms. One such algorithm, used as the default in BERTopic, is HDBSCAN \cite{mcinnes2017hdbscan}. HDBSCAN simplifies the task by preventing the need to manually select the number of clusters and excelling at automatically clustering dense areas of varying densities. To harness GPU acceleration benefits, we opted for cuML's HDBSCAN implementation. 


\item \textbf{Seed Topics:} 
Different variations of topic modeling are supported by BERTopic, such as dynamic, multimodal, hierarchical, online, and semisupervised approaches. Using vanilla BERTopic, the generated topics may be biased toward terms that are not relevant to the prompt bias mistake categories. To address this issue and guide the topic model towards creating topics that align more closely with our areas of interest, we can introduce a weighting mechanism for specific words within the dataset. This approach ensures that the resulting topic model contains topics whose representations are more in line with our criteria. BERTopic introduces a semisupervised variation of topic modeling known as Guided Topic Modeling, which facilitates this process by allowing us to guide the model's topic creation using weighted terms called \textit{seed}, thereby producing topics that better match our intended criteria. Seed is a nested list of terms, where each inner list consists of phrases relevant to one of our desired topics. The prompt bias mistake categories generated as a result of the method described in subsection ~\ref{subsubsec:Prompt Bias Mistake Categories Generation} were used as the seed topics.


\item \textbf{Mapping Seed Topics:} Mapping our predefined seed topics, which represent prompt bias mistake categories, to the topics generated by the model was a crucial step in assessing the relevance and coverage of our predefined categories within the actual topics identified by the model. This process was essential to ensure that the topics discovered by the model closely aligned with the principles we manually analyzed. To achieve this, we processed each prompt bias mistake category (detailed shown in section~\ref{subsec:prompt bias mistake categories}) into individual words to match BERTopic's output format. Subsequently, we used the \texttt{model.find\_topic} function to identify the most closely matching generated topic, thus establishing a connection between our predefined categories and the topics generated by the model.

\end{enumerate}
After this procedure, the BERTopic model yields the count of each category within the generated high-level rationales. We calculated the prevalence of each category: 
\begin{equation}
\text{Category\_Prevalence} = \frac{category\_count}{\sum generated\_rationales}
\end{equation}

\subsection{Leveraging Prompt Bias Lessons to Improve Model Performance}\label{sec:rq3_method}
To address prompt bias mistakes made by the model during the prediction phase, we propose a method where the model considers these biases when performing the task. Since we don't have access to the model's weights or knowledge of its training data, our approach centers on the prompt itself. Specifically, we suggest that the model can take prompt biases into account by adding a lesson to the prompt. This lesson instructs the model on what to focus on during the task and is referred to as a "prompt bias lesson." In essence, each lesson serves as a high-level encapsulation of the model's confusion, offering valuable insights into potential solutions for rectifying prompt bias mistakes. To generate these prompt bias lessons, two authors conducted an independent review of each prompt bias mistake category and manually assigned corresponding lessons. Following this independent assessment, the two authors discussed the crafted prompt bias lessons. After discussion, both authors reached a unanimous consensus, with an inter-rater reliability Cohen's Kappa (k)~\cite{gwet2008computing} of 1, indicating perfect agreement. Finally, we appended the lesson to the default prompt to create the new prompt.


\subsection{Method Evaluation} Finally, we evaluated the proposed method by rerunning the model again on all datasets with the modified prompts. To illustrate the effectiveness of the model, we report the performance of the model using standard precision, recall, accuracy, and F1 (harmonic mean of precision and recall) because those metrics show the reliability and effectiveness of the model. Moreover, these metrics were used to evaluate the performance of clone detection models in prior work~\cite{wang2020detecting, feng2020codebert}. We then use those metrics to compare the performance of the model before and after adopting the proposed method. Moreover, to evaluate whether the improvement is significant or not, we adopted the t-test~\cite{kim2015t, bland1995multiple, bland1995multiple} and recorded the p-value. We give the definition of each metric below.


\subsubsection{Precision:}
Precision is a measure that quantifies the number of correct positive predictions made. It is defined as the ratio of true positives (tp) to the sum of true positives and false positives (fp). The formula for precision is given by:
\begin{equation}
\text{Precision (P)} = \frac{tp}{tp + fp}
\end{equation}

\subsubsection{Recall:}
Recall, often referred to as the true positive rate, measures the percentage of actual positives that were correctly classified. It is the ratio of true positives to the sum of true positives and false negatives (fn). The formula for recall is:
\begin{equation}
\text{Recall (R)} = \frac{tp}{tp + fn}
\end{equation}

\subsubsection{Accuracy:}
Accuracy is a metric that measures the proportion of true results (both true positives and true negatives) in the dataset. It provides a holistic view of the model's performance across all classes. The formula for accuracy, considering true negatives (tn) as well, is:
\begin{equation}
\text{Accuracy} = \frac{tp + tn}{tp + tn + fp + fn}
\end{equation}

\subsubsection{F1-score:}
The F1-score is the harmonic mean of precision and recall. It provides a balance between the two metrics, especially when there is an uneven class distribution. The formula for the F1-score is:
\begin{equation}
F1 = 2 \times \frac{Precision \times Recall}{Precision + Recall}
\end{equation}


\section{Results}
\label{sec:results}
In this section, we discuss the results of our study by placing them in the context of three research questions, which are defining the prompt bias mistake categories (RQ1), finding the prevalence of each prompt bias mistake category and the most frequent category (RQ2), and whether we can improve the model performance on clone detection task by leveraging prompt bias mistake categories (RQ3).

\subsection{RQ1: In the context of within-language and cross-language clone detection tasks, what are the prompt bias mistake categories?}
In this section, we present the models' selection results, the default prompt we generated from GPT-4, the selected model's baseline performance, the confidence level of the model's incorrect predictions, and prompt bias mistake categories.

\begin{table*}[]
\footnotesize
\centering
\caption{Models Selection Result}
\label{tab:model-selection_result}
\resizebox{0.7\columnwidth}{!}{%
\begin{tabular}{|l|l|c|c|c|c|}
\hline
\textbf{Model} & \textbf{Benchmark Dataset Name} & \textbf{Precision} & \textbf{Accuracy} & \textbf{Recall} & \textbf{F1 Score} \\ \hline
\multirow{4}{*}{\centering Turbo-3.5} & avatar without comment & 1.00 & 0.66 & 0.32 & 0.48 \\
& avatar with comment & 1.00 & 0.58 & 0.17 & 0.29 \\
& poolC with comment & 0.96 & 0.91 & 0.87 & \textbf{0.91} \\ 
& poolC without comment & 0.96 & 0.84 & 0.72 & 0.82 \\ \hline
\multirow{4}{*}{\centering Text\_002} & avatar without comment & 0.51 & 0.53 & 1.00 & 0.68 \\
& avatar with comment & 0.51 & 0.52 & 1.00 & 0.68 \\
& poolC with comment & 0.53 & 0.53 & 0.99 & 0.69 \\ 
& poolC without comment & 0.54 & 0.55 & 1.00 & 0.70 \\ \hline
\multirow{4}{*}{\centering PaLM} & avatar without comment & 0.96 & 0.92 & 0.88 & \textbf{0.92} \\
& avatar with comment & 0.99 & 0.87 & 0.75 & \textbf{0.86} \\
& poolC with comment & 1.00 & 0.90 & 0.81 & 0.89 \\
& poolC without comment & 1.00 & 0.85 & 0.71 & \textbf{0.83} \\ \hline
\multirow{4}{*}{\centering CodeT5} & avatar without comment & 0.55 & 0.54 & 0.57 & 0.56 \\
& avatar with comment & 0.49 & 0.50 & 0.56 & 0.52 \\
& poolC with comment & 0.47 & 0.0.45 & 0.45 & 0.46 \\
& poolC without comment & 0.58 & 0.56 & 0.62 & 0.60 \\ \hline
\end{tabular}%
}
\end{table*}

\subsubsection{Model Selection Results:}\label{subsubsec:model_selection_results} We use the default prompt shown in section~\ref{subsubsec: Selecting Model} to select the highest-performing model and to establish the model's baseline performance.

Subsequently, we present each model's performance across all benchmark datasets in Table \ref{tab:model-selection_result}. Upon a meticulous review of these results, a consistent trend emerged, with the PaLM model consistently surpassing the other models across nearly all benchmark datasets. The only exception to this pattern was observed in the case of the poolC with comment benchmark dataset, where we observed a marginal 2\% higher F1 score for the Turbo-3.5 model compared to the PaLM model. Nonetheless, considering that the PaLM model outperformed the Turbo-3.5 model across all other instances and that the 2\% discrepancy in F1 score is relatively modest, we ultimately selected the PaLM model as our primary choice for further experimentation.

\begin{table*}[]
\footnotesize
\centering
\caption{Confidence Level Of PaLM Models}
\label{tab:confidence level results}
\resizebox{\columnwidth}{!}{%
\begin{tabular}{|c|c|c|c|c|}
\hline
Confidence Level & poolC with comment frequency & poolC without comment frequency & avatar with comment frequency & avatar without comment frequency \\ \hline
0 & 100.00\% & 99.72\% & 76.33\% & 86.44\% \\ \hline
60 & 0.00\% & 0.05\% & 0.00\% & 0.00\% \\ \hline
80 & 0.00\% & 0.05\% & 6.28\% & 0.00\% \\ \hline
90 & 0.00\% & 0.19\% & 15.94\% & 6.78\%  \\ \hline
95 & 0.00\% & 0.00\% & 0.00\% & 6.78\% \\ \hline
99 & 0.00\% & 0.00\% & 0.48\% & 0.00\% \\ \hline
100 & 0.00\% & 0.00\% & 0.97\% & 0.00\% \\ \hline
\end{tabular}%
}
\end{table*}

Lastly, we present the confidence level of the PaLM model for all incorrect predictions in both the poolC and avatar datasets, as detailed in Table ~\ref{tab:confidence level results}. Upon a comprehensive analysis of these results, a clear and consistent pattern emerges. Specifically, when the confidence level equals or exceeds 80, the PaLM model demonstrates a notably high level of confidence in its predictions. This is evident from the majority of cases where the confidence level consistently remains above 80 or precisely at 80. It is worth noting that there are rare instances where the confidence level drops to 60. However, this occurrence is limited to the poolC without comment dataset and is exceptionally infrequent, constituting only 0.05\% of all instances. As a result, we can confidently affirm that the PaLM model demonstrates a robust level of confidence in its predictions when its confidence level reaches or exceeds 80. This finding serves as strong validation for our decision to employ 80 as the threshold to identify reliable incorrect predictions, as discussed in section ~\ref{subsec:confidence}.


\begin{figure}[t]
    \centering
    \begin{minipage}[b]{0.4\textwidth}
        \centering
        \includegraphics[width=\textwidth]{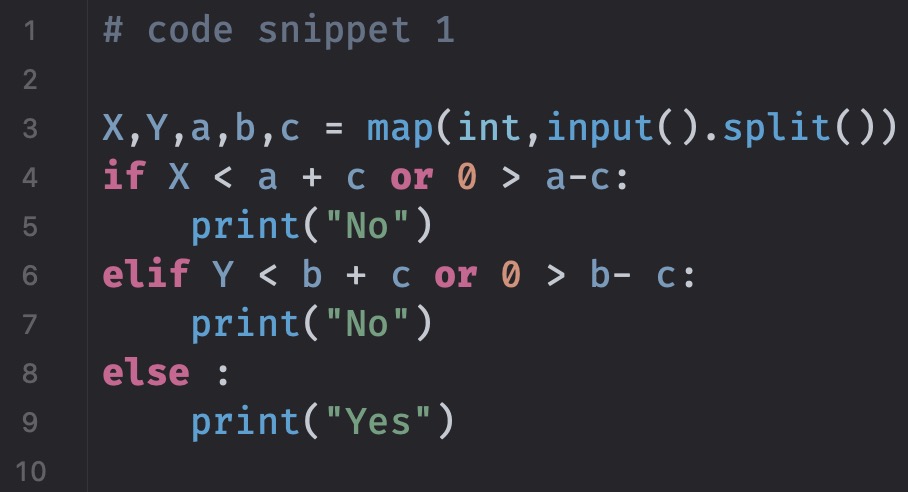}
    \end{minipage}
    \hfill
    \begin{minipage}[b]{0.5\textwidth}
        \centering
        \includegraphics[width=\textwidth]{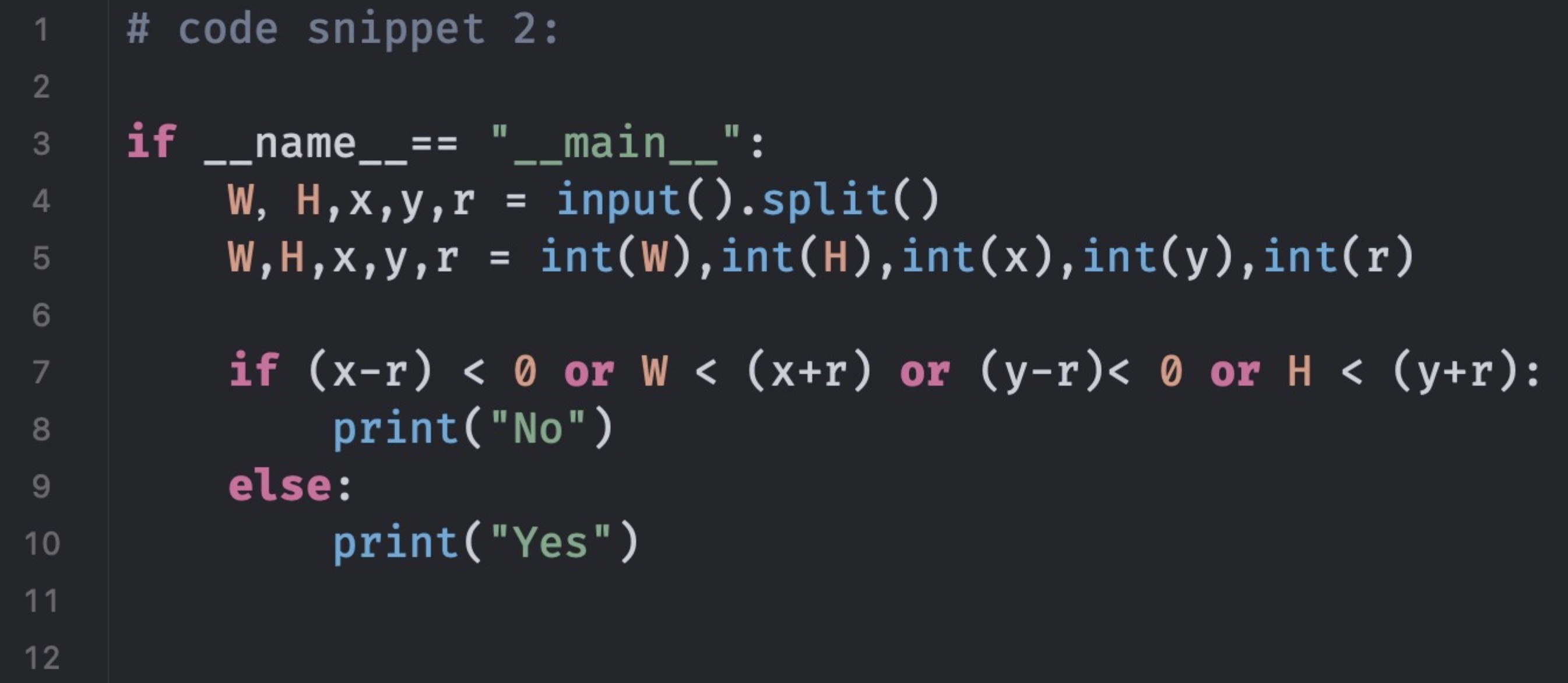}
    \end{minipage}
    \caption{Example of Operational Ambiguities category}
    \label{fig:operational_examples}
\end{figure}
\subsubsection{Prompt Bias Mistake Categories:}\label{subsec:prompt bias mistake categories} Next, we present our defined eight prompt bias mistake categories along with their definition:

\begin{itemize} 

\item \textit{Misconception Of Clone Code Definition:} This category refers to LLMs' confusion regarding the definition of clone code. Their interpretation of clone code aligns precisely with Type 1 clone code, which denotes code snippets that are identical~\cite{roy2007survey}. Consequently, LLMs categorize all clone codes falling under Type 2-4 as non-clone codes~\cite{roy2007survey}.

\item \textit{Operational Ambiguities:} This category pertains to operators. LLMs tend to classify code snippets as non-clone codes when they exhibit variations in the order of operations. Additionally, LLMs often misclassify code snippets that differ in statements containing semantically equivalent but textually distinct operators as non-clone codes. In both cases related to operations, these code snippets are semantically identical. An illustrative example is provided in Figure~\ref{fig:operational_examples}, where variables X, Y, a, b, c are equated to W, H, x, y, r, respectively. LLMs incorrectly categorize this instance as non-clone code due to their confusion arising from the textual differences between "(x-r) < 0" and "0 > a-c," where "a=x" and "r=c." In essence, when LLMs make mistakes in this category, their ability to identify clone codes becomes limited to Type 2 clones~\cite{roy2007survey}.


\begin{figure}[t]
    \centering
    \begin{minipage}[b]{0.45\textwidth}
        \centering
        \includegraphics[width=\textwidth]{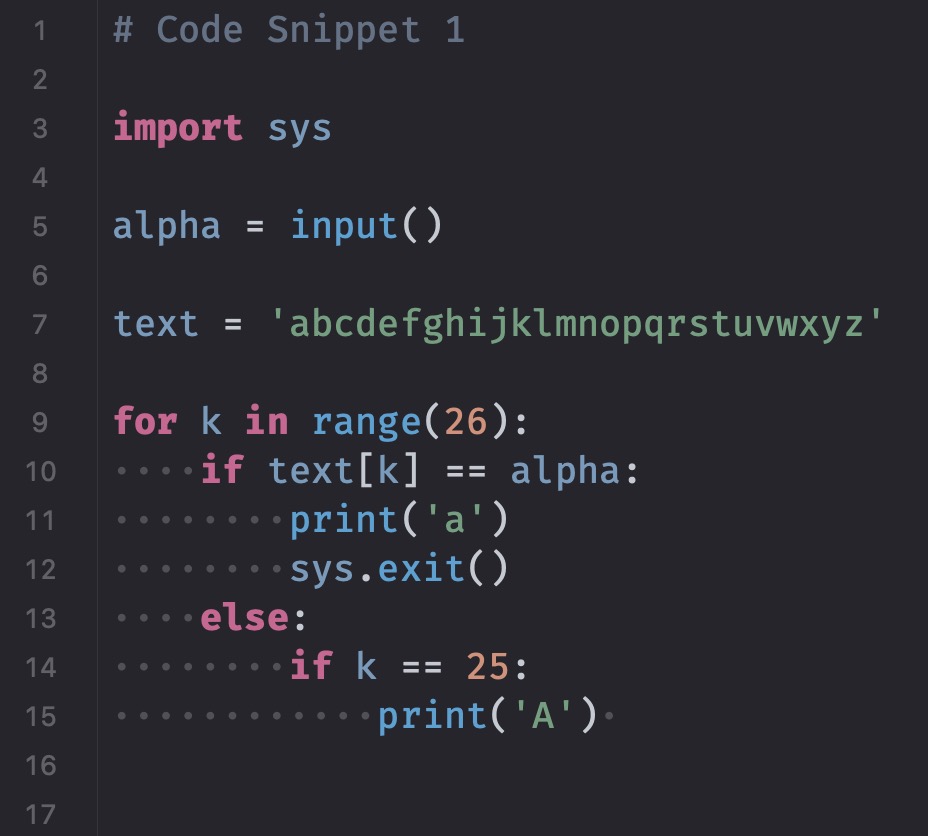}
    \end{minipage}
    \hfill
    \begin{minipage}[b]{0.52\textwidth}
        \centering
        \includegraphics[width=\textwidth]{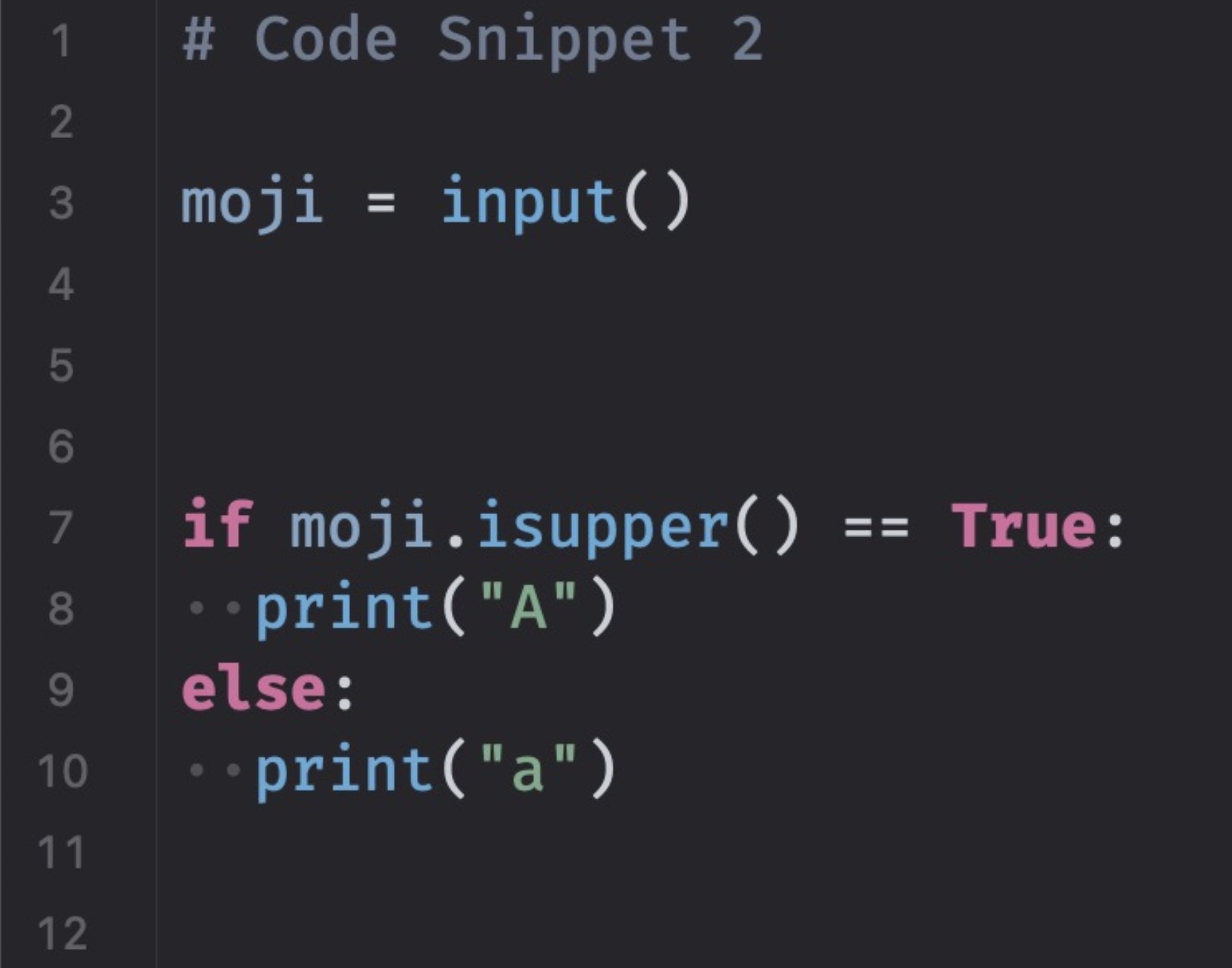}
    \end{minipage}
    \caption{Example of Confusion Over Varied Approaches category}
    \label{fig:approach_examples}
\end{figure}

\item \textit{Variable Initialization And Naming Ambiguity:} This category pertains to variables, encompassing aspects such as the sequence of variable initialization and variable names. Regarding the order of variable initialization, LLMs often categorize code snippets as non-clone codes when they differ in the sequence of variable initialization. Conversely, in situations where two code snippets employ distinct variable names, but the roles and values of these variables in the source code remain unchanged, LLMs inaccurately classify them as non-clone codes. In sum, when LLMs make errors within this category, their capability to detect clone codes is confined to Type 1 and 2 clones~\cite{roy2007survey}.

\item \textit{Data Structure-Based Misclassification:} This category arises when two code snippets differ in their choice of data structures, a divergence that does not impact the underlying logic of the code snippets. For instance, when code snippet 1 utilizes a list to store elements, while code snippet 2 employs a set for the same purpose. LLMs often make mistakes by inaccurately categorizing such code snippets as non-clone codes. In essence, when LLMs make mistakes within this category, their capacity to identify clone codes is constrained primarily to Type 1 and Type 2 clones~\cite{roy2007survey}.

\item \textit{Misinterpretation Of Function And Library API Nomenclature:} This category pertains to the nomenclature of functions, encompassing both user-defined functions and APIs sourced from libraries. Within code snippets, it's common to encounter user-defined functions that may possess dissimilar function names but exhibit identical or akin functionalities. Conversely, APIs originating from distinct libraries often share similar functionalities. As an illustration, consider the functions torch.matmul() from the PyTorch library and np.matmul() from the NumPy library, which serve the same purpose despite having different names. LLMs make erroneous classifications when confronted with such code snippets. In essence, when LLMs err in this category, their capacity to identify clone codes becomes confined to Type 1 clones~\cite{roy2007survey}.

\item \textit{Thematic Distraction In Semantics:} This category pertains to situations in which two code snippets perform semantically identical tasks, yet they differ in their thematic content. For instance, consider code snippet 1, which employs binary search to locate a specific card within a poker card deck, and code snippet 2, which utilizes binary search to find a number within a list of numbers. If we disregard the thematic variations between these two code snippets, they essentially accomplish the same task. However, LLMs often become perplexed by the thematic distinctions and consequently misclassify them as non-clone code instances. As a result, when LLMs make mistakes in this category, their capability to identify clone codes is limited to Type 1, 2, and 3 clones~\cite{roy2007survey}.

\item \textit{Erroneous Code Comprehension:} This category pertains to instances where LLMs encounter challenges in comprehending the code snippet or misconstrue its context. When faced with such situations, the LLMs begin producing arbitrary explanations for the code snippet, akin to the issue of hallucination as described by Lee et al.\cite{lee2023mathematical}. Consequently, when LLMs make mistakes within this category, their capacity to detect clone codes is confined primarily to Type 1 clones~\cite{roy2007survey}.

\item \textit{Confusion Over Varied Approaches:} This classification pertains to scenarios in which two code snippets exhibit comparable logic but differ in their textual representation, especially when two code snippets adopt slightly different approaches. A clone code example of this category is shown in Figure~\ref{fig:approach_examples}. Both code snippets essentially perform the same function: they print 'a' if the input value is a lowercase letter and 'A' if the input value is an uppercase letter. However, PaLM classified them as non-clone code due to their textual dissimilarity. LLMs often overlook the similarity in code logic between these two snippets and instead make predictions primarily based on the textual variations. In essence, when LLMs err in this category, their capability to discern clone codes is limited to Type 1, 2, and 3 clones~\cite{roy2007survey}.
\end{itemize}

\begin{tcolorbox}
Finding 1: Majority of the mistakes made by LLM limits its capability to identify Type 4 clones.
\end{tcolorbox}
\subsection{RQ2: How prevalent are the mistake categories and what is the most frequent category?} \label{subsection: RQ2}

In this section, we provide an overview of the frequency of each prompt bias mistake category drawn from BERTopic model's output. For better illustration purposes, we average the frequency of each category across with and without comment versions of poolC and avatar datasets. We present the results in Table ~\ref{tab:rq2_results}. From the result, we observed that the category "Misinterpretation Of Function and Library API Nomenclature" has high frequency in both poolC and avatar datasets indicating that it is the most prevalent category (65.39\%) in all datasets.


Moreover, when we contrast the avatar dataset with the poolC dataset, we observe a significant increase in frequency values across all categories. This variance can be attributed to the computation method employed for determining frequency values, which takes into account the number of topics generated by the BERTopic model from the incorrect predictions. Remarkably, the PaLM model displayed outstanding performance when applied to the avatar dataset. Consequently, the BERTopic model could only produce three topics due to small number of wrong predictions. As a result, this led to the merging of several categories into the same topics, thereby causing the observed rise in frequency values. 


\begin{tcolorbox}
Finding 2: Misinterpretation Of Function and Library API Nomenclature is the most frequent prompt bias mistake category found in within and cross language clone detection datasets.
\end{tcolorbox}

\begin{table*}[]
\footnotesize
\centering
\caption{Frequency of Each Category}
\label{tab:rq2_results}
\resizebox{\columnwidth}{!}{%
\begin{tabular}{|l|c|c|c|}
\hline
\multicolumn{1}{|c|}{Prompt Bias Mistake Category Name} & \multicolumn{1}{c|}{\textbf{poolC}} & \multicolumn{1}{c|}{\textbf{avatar}} & \multicolumn{1}{c|}{\textbf{Across Datasets}} \\ \cline{2-4}
& Frequency (in\%) & Frequency (in \%) & Average Frequency (in \%)\\ \hline
Misinterpretation Of Function and Library API Nomenclature & \textbf{36.38\%} & \textbf{94.40\%} & \textbf{65.39\%} \\ \hline
Variable Initialization And Naming Ambiguity & 35.42\% & 82.09\% & 58.76\% \\ \hline
Operational Ambiguities & 29.11\% & 48.13\% & 38.62\% \\ \hline
Misconception Of Clone Code Definition & 27.14\% & 48.13\% & 37.64\%  \\ \hline
Erroneous Code Comprehension & 27.24\% & 42.91\% & 35.08\% \\ \hline
Overemphasis On Textual Similarity & 9.12\% & 60.45\% & 34.79\%\\ \hline
Data Structure-based Misclassification & 1.28\% & 47.39\% & 24.34\% \\ \hline
Thematic Distraction In Semantics & 8.70\% & 48.13\% & 28.42\% \\ \hline
\end{tabular}%
}
\end{table*}


\begin{table}[htbp]
\footnotesize
\centering
\caption{Overview Of Prompt Bias Lessons}
\label{tab:prompt bias lessons}
\resizebox{0.8\columnwidth}{!}{%
\begin{tabular}{|c|p{10cm}|}
\hline
\textbf{Lesson ID} & \textbf{Prompt Bias Lessons}\\
\hline
1 & Clone code is defined by similar functionalities; therefore, the code does not necessarily need to be identical. \\ \hline
2 & Differences in function, and method names should not be criteria for determining whether code is cloned. \\ \hline
3 & Differences in variable names should not be criteria for determining whether code is cloned. \\ \hline
4 & Variations in approach should not be criteria for determining whether code is cloned. \\ \hline
5 & Differences in data structure and implementation details should not be criteria for determining whether code is cloned. \\ \hline
6 & Code syntax should not be a predominant factor in determining clones. \\ \hline
7 & Code logic holds greater significance than minor code differences in determining whether code is cloned. \\ \hline
8 & Variations in the order of operations should not be used to determine code clones. \\ 
\hline
\end{tabular}
}
\label{tab:false-connections}
\end{table}

\subsection{RQ3: How can we harness these categories to improve the model performance?}
In this section, we present our findings pertaining to the response to RQ3. Initially, we showcase the baseline performance of the PaLM model on each dataset using the default prompt shown in section~\ref{subsubsec:model_selection_results}, in Tables ~\ref{tab:avatar-prompt-bias-lesson-performances} and ~\ref{tab:pool-prompt-bias-lesson-performances} indicated as ``deafult''. Subsequently, we provide an overview of the prompt bias lessons that were developed based on the categories of prompt bias mistakes, as detailed in Table ~\ref{tab:prompt bias lessons}. The two authors performed qualitative analysis~\cite{strauss1998basics} to derive the lessons from prompt bias mistake categories. Details were described in section~\ref{sec:rq3_method}.



To assess the effectiveness of our proposed approach, we carried out an ablation study~\cite{meyes2019ablation} to investigate the impact of individual prompt bias lesson on the performance of the PaLM model. The outcomes for both the poolC and avatar datasets are summarized in Table~\ref{tab:avatar-prompt-bias-lesson-performances} and Table~\ref{tab:pool-prompt-bias-lesson-performances}. To facilitate a clearer understanding of the improvement brought about by each lesson, both tables contain a column labeled $\Delta_{F1}$, indicating the change in the model's F1 score. A "/" denotes instances where no discernible improvement was achieved from a particular lesson.

Upon analyzing the $\Delta_{F1}$ column, we observed an average improvement of approximately \textit{3.99\%} for the avatar without comment dataset and \textit{5.16\%} for the avatar with comment dataset. For the poolC dataset, the average improvement was \textit{3.25\%} for without comment dataset and \textit{3.03\%} for with comment dataset. Additionally, observing from the $\Delta_{F1}$ columns in both Table~\ref{tab:avatar-prompt-bias-lesson-performances} and Table~\ref{tab:pool-prompt-bias-lesson-performances}, lesson 7 consistently brings substantial improvements to the performance of the PaLM model in all four datasets, indicating that lesson "Code logic holds greater significance than minor code differences in determining whether code is cloned." is the most effective lesson.

To determine the maximum performance boost achievable with our proposed method, we ran the model with all lessons incorporated into the default prompt. The results are also displayed in Table~\ref{tab:avatar-prompt-bias-lesson-performances} and Table~\ref{tab:pool-prompt-bias-lesson-performances}. Notably, our analysis showed that the peak improvements of our methods can be up to \textit{10.81\%} in terms of F1 score for the avatar dataset and \textit{9.77\%} for the poolC dataset.

Furthermore, we conducted a t-test to assess the statistical significance of these improvements. The reported p-values shown in Table~\ref{tab:pool-prompt-bias-lesson-performances} and Table~\ref{tab:avatar-prompt-bias-lesson-performances} demonstrated that all improvements were statistically significant, as they were below the threshold of 0.05~\cite{halsey2015fickle}. In summary, our proposed approach demonstrates its effectiveness and validity through the significant enhancements achieved exclusively through the incorporation of additional lessons. Figure~\ref{pic:lessons_all_in_prompt} shows the prompt that resulted in the best performance.

\begin{tcolorbox}
Finding 3: Our proposed method is effective and brings an improvement up to 10.81\% and 9.77\% in terms of F1 score for avatar and poolC datasets, respectively.
\end{tcolorbox}

\begin{tcolorbox}
Finding 4: "Code logic holds greater significance than minor code differences in determining whether code is cloned." is the most effective individual lesson.
\end{tcolorbox}

\begin{figure}[h]
\begin{tcolorbox}
\textit{You are a helpful assistant that takes two pieces of code snippets and detects whether these two pieces of code snippets are clone code or not. The first code snippet is written in <language1>, and the second code snippet is written in <language2>. <prompt bias lessons 1>. <prompt bias lessons 2>. <prompt bias lessons 3>. <prompt bias lessons 4>. <prompt bias lessons 5>. <prompt bias lessons 6>. <prompt bias lessons 7>. <prompt bias lessons 8>. Here are the two code snippets: Code Snippet1: <code1>. Code Snippet2: <code2>. If you think any part of the two code snippets contains clones,
answer Yes; otherwise answer No. You must answer either Yes or No. Your Response:[INSERT].}
\end{tcolorbox}
\caption{Best Performing Prompt}
\label{pic:lessons_all_in_prompt}
\end{figure}

Moreover, in order to gain a deeper understanding of the effectiveness of our approach, we conducted a comparative analysis between the predictions made by the PaLM model using the default prompt and those using the prompt enriched with all\_lessons (as illustrated in Figure ~\ref{pic:lessons_all_in_prompt}). The results of this analysis are presented in Table~\ref{tab:prediction_conversion}. To be specific, we quantified the number of instances that were initially misclassified by the PaLM model when using the default prompt but were subsequently correctly classified when employing the prompt containing all\_lessons (column 2). Conversely, we calculated the number of instances that were initially correctly classified by the PaLM model when using the default prompt but were subsequently incorrectly classified when employing the prompt containing all\_lessons (column 3). 

\begin{table}[h!]
\footnotesize
\centering
\caption{Prediction Shifts for PoolC and Avatar: Default Prompt vs. With All Lessons Prompt}
\begin{tabular}{|c|c|c|c|}
\hline
\textbf{Dataset Name} & \textbf{Incorrect to Correct Predictions Count} & \textbf{Correct To Incorrect Predictions Count} \\
\hline
poolC without comment & 1,356 & 77 \\
\hline
poolC with comment & 1,166 & 101 \\
\hline
avatar without comment & 128 & 25 \\
\hline
avatar with comment & 175 & 20 \\
\hline
\end{tabular}
\label{tab:prediction_conversion}
\end{table}
Our findings indicate that through the adoption of our proposed method, the PaLM model makes a greater number of correct predictions, as evident from the significant values in column 2. Importantly, this enhanced performance does not negatively affect the previously correct predictions, as reflected in the small values in column 3. 

\begin{table*}[]
\footnotesize
\centering
\caption{The Performances of PaLM Model With Prompt Bias Lesson on Avatar Dataset}
\label{tab:avatar-prompt-bias-lesson-performances}
\resizebox{\textwidth}{!}{
\begin{tabular}{|l|c|c|c|c|c|c|c|c|c|c|c|c|}
\hline
\multicolumn{1}{|c|}{Prompt Lesson ID} & \multicolumn{6}{c|}{\textbf{avatar without comment}} & \multicolumn{6}{c|}{\textbf{avatar with comment}} \\ \cline{2-13}
& Precision & Accuracy & Recall & F1 score & $ \Delta_{F1} (in \%)$ & P-value & Precision & Accuracy & Recall & F1 score & $ \Delta_{F1} (in \%)$ & P-value \\ \hline
1 & 97.10 & 93.72 & 90.36 & 93.61 & 4.31\% & <5.76e-12* & 98.03 & 93.36 & 88.63 & 93.09 & 6.94\% & <1.76e-20* \\
2 & 97.12 & 93.33 & 89.49 & 93.15 & 3.85\% & <2.29e-13* & 98.17 & 92.67 & 87.14 & 92.33 & 6.18\% & <7.12e-18* \\
3 & 97.45 & 92.97 & 88.43 & 92.72 & 3.42\% & <5.29e-13* & 98.14 & 91.91 & 85.65 & 91.47 & 5.32\% & <4.95e-14* \\
4 & 98.41 & 86.08 & 73.54 & 84.18 & / & / & 98.16 & 84.37 & 70.17 & 81.84 & / & / \\
5 & 97.93 & 90.73 & 83.55 & 90.17 & 0.87\% & <1.48e-03* & 98.25 & 88.99 & 79.62 & 87.96 & 1.81\% & <1.31e-06* \\
6 & 97.92 & 90.86 & 83.77 & 90.29 & 0.99\% & <7.73e-03* & 98.42 & 89.65 & 80.85 & 88.77 & 2.62\% & <3.89e-09* \\
7 & 97.36 & 94.85 & 92.30 & \textbf{94.77} & \textbf{5.47\%} & <2.09e-21* & 98.08 & 94.39 & 90.66 & \textbf{94.23} & \textbf{8.08\%} & <2.78e-26* \\
8 & 98.49 & 88.33 & 78.04 & 87.08 & / & / & 98.45 & 87.27 & 75.87 & 85.70 & / & / \\ \hline
all\_lessons & 95.48 & 96.39 & 97.51 & \textbf{96.48} & \textbf{7.18\%} & <1.65e-27* & 96.28 & 96.92 & 97.64 & \textbf{96.96} & \textbf{10.81}\% & <3.39e-27* \\ \hline
default & 98.20 & 90.08 & 82.01 & 89.30 & 0.00 & / & 98.18 & 87.53 & 76.75 & 86.15 & 0 & / \\
\hline
\end{tabular}%
}
\end{table*}

\begin{table*}[]
\footnotesize
\centering
\caption{The Performances of PaLM Model With Prompt Bias Lesson On PoolC Dataset}
\label{tab:pool-prompt-bias-lesson-performances}
\resizebox{\textwidth}{!}{
\begin{tabular}{|l|c|c|c|c|c|c|c|c|c|c|c|c|}
\hline
\multicolumn{1}{|c|}{Prompt Lesson ID} & \multicolumn{6}{c|}{\textbf{poolC without comment}} & \multicolumn{6}{c|}{\textbf{poolC with comment}} \\ \cline{2-13}
& Precision & Accuracy & Recall & F1 score & $ \Delta_{F1}$ (in \%) & P-value & Precision & Accuracy & Recall & F1 score & $ \Delta_{F1}$ (in\%) & P-value \\ \hline
1 & 98.91 & 90.76 & 82.23 & 89.81 & 5.27\% & <2.13e-17* & 98.54 & 91.64 & 84.48 & 90.97 & 4.56\% & <1.26e-12* \\
2 & 99.03 & 87.85 & 76.30 & 86.19 & 1.65\% & <3.44e-05* & 98.78 & 89.21 & 79.26 & 87.95 & 1.54\% & <8.38e-05* \\
3 & 99.06 & 87.13 & 74.81 & 85.25 & 0.71\% & <9.02e-03* & 98.75 & 88.76 & 78.38 & 87.39 & 0.98\% & <7.41e-07* \\
4 & 99.50 & 78.95 & 57.84 & 73.16 & / & / & 99.21 & 82.11 & 64.44 & 78.13 & / & / \\
5 & 99.45 & 83.26 & 66.61 & 79.78 & / & / & 99.23 & 85.55 & 71.45 & 83.08 & / & / \\
6 & 99.51 & 81.17 & 62.44 & 76.73 & / & / & 99.13 & 84.31 & 68.97 & 81.35 & / & / \\
7 & 98.92 & 90.81 & 82.37 & \textbf{89.89} & \textbf{5.35\%} & <5.30e-17* & 98.54 & 92.06 & 85.28 & \textbf{91.43} & \textbf{5.02\%} & <5.06e-23* \\
8 & 99.70 & 78.88 & 57.67 & 73.07 & / & / & 99.48 & 82.17 & 64.41 & 78.20 & / & / \\ \hline
all\_lessons & 98.26 & 94.57 & 90.67 & \textbf{94.31} & \textbf{9.77\%} & <9.23e-34* & 97.77 & 94.75 & 91.58 & \textbf{94.55} & \textbf{8.14\%} & <8.70e-24* \\ \hline
default & 98.96 & 86.59 & 73.79 & 84.54 & 0 & / & 98.62 & 87.97 & 76.89 & 86.41 & 0 & / \\
\hline
\end{tabular}%
}
\end{table*}

\section{Discussion and Implications}
\label{sec:discussion}
This section presents the implications of our findings for researchers and practitioners who consider using LLMs in their work.

\vspace{0.1cm}
\textbf{Implications for practitioners:} In comparison to BERT models, LLMs exhibit impressive source code comprehension abilities, making them a well-suited choice for the task of clone detection. However, practitioners who are either contemplating the adoption of LLMs or are currently utilizing them need to be aware of the notable performance variability of LLM and implement the lessons outlined in Table~\ref{tab:prompt bias lessons} to mitigate prompt bias. 



\vspace{0.1cm}
\textbf{Implications for researchers:} In this study, we conducted a comprehensive comparison between CodeT5, a high-performing BERT model, and several popular GPT models within the context of clone detection. Our findings revealed that GPT models consistently outperformed CodeT5, underscoring the superiority of GPT models in this domain. Therefore, we strongly advocate for increased attention from researchers toward GPT models as a result of these impressive performances. Additionally, our research sheds light on the issue of prompt bias in GPT models and introduces a methodology involving prompt bias lessons to alleviate this problem. We encourage researchers to prioritize investigations into this specific concern related to GPT models and develop further solutions for mitigating prompt bias. Furthermore, we recommend clone detection researchers consider incorporating LLMs into their work as our work has shown the supreme performance of LLMs over other machine learning models. 

As our work have shown, erroneous code comprehension is one potential reason for LLMs' incorrect predictions. Hence, future research should concentrate on addressing this issue. A promising avenue for exploration involves assisting LLMs in their comprehension through prompting or providing them with explicit explanations of the source code. 
Furthermore, we manually generate prompt bias mistake categories and lessons. However, it would be more efficient and pragmatic if future research could automate this step. Lastly, as we have shown that prompt bias lessons deriving from prompt bias mistake categories have the potential to mitigate model prompt bias, an intriguing avenue for future research would involve uncovering additional categories and devising supplementary prompt bias lessons.

Moreover, within this work, we conducted an ablation study to assess the performance enhancements gained from individual lessons and calculated the cumulative performance improvement when applying all lessons simultaneously. However, we posit that there may exist optimal combinations of lessons that could yield even better results. Therefore, we encourage researchers to explore and identify more efficient ways of determining the most effective lesson combination.

\section{Threats to Validity}


\label{sec:ttv}
In this study, our pipeline encompasses several key steps: model selection, the generation of rationales for selected model's incorrect predictions, defining categories for prompt bias mistakes, crafting prompt bias lessons, and enhancing the selected model through modified prompts. We have taken meticulous measures to mitigate potential threats that might affect the validity of our research. Nevertheless, it is important to acknowledge the possibility of flaws in our pipeline.

During the model selection phase, we observed outstanding performance from GPT models on both the poolC and avatar datasets. However, it's essential to note that this performance may not necessarily extend to clone detection datasets written in different programming languages.

In the process of defining prompt bias mistake categories, our approach mainly relied on the model's incorrect prediction groups within the poolC dataset, with a limited sample size of 100 instances. Consequently, the prompt bias mistake categories we identified may be confined to the characteristics specific to the source code within the poolC dataset, and they may not encompass all potential prompt bias mistakes.

Furthermore, while applying the prompt bias lessons led to notable improvements of 10.81\% and 9.77\% for the avatar and poolC datasets, respectively, it's important to exercise caution when generalizing these findings to other clone detection datasets. However, we contend that our approach still provides valuable insights into mitigating prompt bias in large language models as we have shown the generalizability of the method in avatar dataset.

Lastly, the definition of prompt bias mistake categories and the generation of prompt bias lessons were conducted manually. To minimize the introduction of unintentional bias, both authors independently reviewed the model-generated mistakes, formulated the prompt bias categories, and crafted the prompt bias lessons. The high Cohen's kappa of 0.9 indicates a high level of reliability in this process.

\section{Conclusion}
\label{conclusion}
Clone detection has remained a longstanding and prominent research domain due to the persistence of clone codes in traditional software development, where developers frequently employ code snippets from online sources. The advent of LLMs, notably exemplified by ChatGPT, has brought code generation into the forefront of their capabilities. However, this capability has the potential to lead to an increase in clone code instances, as LLMs generate code patterns they've encountered during their training. Consequently, the significance of clone detection has become more pronounced than ever.

In this study, we explored the feasibility and suitability of utilizing LLMs for clone detection and found that LLMs are well-suited for this task. Additionally, we delved into prompt bias in LLMs and introduced a comprehensive framework for its mitigation. Our findings illuminated the fact that although LLMs perform admirably in clone detection, they are not infallible. Our analysis unveiled categories of prompt bias mistake that underscore some of the weaknesses in LLMs, including misconception of clone code definition, ambiguities related to the order of operations and semantically equivalent operations, uncertainties in variable initialization and naming, misclassification based on data structures, misinterpretations of function and API nomenclature, thematic distraction in semantics, erroneous code comprehension, and confusion over varied Approaches. Our proposed method shows possible directions for improving the weakness of the LLMs.

These findings offer valuable insights for researchers and practitioners considering the integration of LLMs into their work. As we chart the course for the future of software engineering, we posit that fully harnessing the potential of LLMs in diverse fields will be a prevailing trend and our work takes a significant first step towards that direction.

\section{Data Availability}
As part of our commitment to open science policy, all data collected for this study are made available as supplemental material. We provide our replication package in~\cite{replication}.
\bibliography{acmart}
\bibliographystyle{IEEEtranS}

\end{document}